\documentclass[useAMS,usenatbib]{mnras}
\usepackage{natbib}
\usepackage{bm}
\usepackage[utf8]{inputenc}
\usepackage{graphicx}	% Including figure files
\usepackage{amsmath}	% Advanced maths commands
\usepackage{amssymb}	% Extra maths symbols
\usepackage{txfonts}
\usepackage{xspace}
\usepackage{rotate} 
\usepackage{subfigure}
\usepackage{float}

\def\apj{ApJ}
\def\apjl{ApJL}
\def\apjs{ApJ Suppl. Ser.}

\def\aap{A\&A}

\def\mnras{MNRAS}

\def\beq#1{\begin{equation}\label{#1}}
\def\eeq{\end{equation}}
\def\beqa#1{\begin{eqnarray}\label{#1}}
\def\eeqa{\end{eqnarray}}

\def\comment#1{\relax}

\def\src{V1239~Her}

\newcommand{\msun}{\mathrm{M}_\odot}
\newcommand{\rsun}{\mathrm{R}_\odot}

\newcommand{\Ry}{\mathrm{Ry}}

\newcommand{\vs}{v_\mathrm{s}}
\newcommand{\di}{\mathrm{d}}

\title[3D accretion disc in V1239 Her]{3D modelling of accretion disc in eclipsing binary system V1239 Her}
\author[Lukin et al.] {
\parbox{\textwidth}{\raggedright V.V.~Lukin$^{1,2}$\thanks{E-mail: vvlukin@gmail.com}, K.L.~Malanchev$^{3,4}$, N.I.~Shakura$^{3}$, K.A.~Postnov$^{3,5}$, 
V.M.~Chechetkin$^{1,6}$, V.P.~Utrobin$^{5}$
}
\\
\\
$^{1}$ M.V. Keldysh Institute of Applied Mathematics RAS, Miusskaya sq., 4, Moscow 125047, Russia\\
$^{2}$ Bauman Moscow State Technical University, 2nd Baumanskaya st., 5, Moscow 105005, Russia \\
$^{3}$ Sternberg Astronomical Institute, Moscow M.V. Lomonosov State University, Universitetskij pr., 13,  Moscow 119234, Russia\\
$^{4}$ Faculty of Physics, M.V. Lomonosov Moscow State University, Leninskie Gory, 1, Moscow 119991, Russia\\
$^{5}$ Institute of Theoretical and Experimental Physics, Bolshaya Cheremushinskaya st., 25, Moscow 117218, Russia\\
$^{6}$ Kurchatov Institute National Research Center, Academician Kurchatov sq., 1, Moscow 123098, Russia}
\begin{document}

\date{Received ... Accepted ...}
\pagerange{\pageref{firstpage}--\pageref{lastpage}} \pubyear{2017}

\maketitle

\label{firstpage}
%\onecolumn
\begin{abstract}
We present the results of 3D-hydrodynamical simulations of accretion flow in 
the eclipsing dwarf nova V1239 Her in quiescence. The model includes the optical star filling its Roche lobe, 
a gas stream emanating from the inner Lagrangian point of the binary system, and the accretion disc
structure. A cold hydrogen gas stream is initially emitted towards a point-like gravitational centre.
A stationary accretion disc is formed in about 15 orbital periods after the beginning
of accretion. The model takes into account partial ionization of hydrogen and
uses realistic cooling function for hydrogen. 
The light curve of the system is calculated as the volume emission of optically thin 
layers along the line of sight up to the optical depth $\tau=2/3$ calculated using Planck-averaged opacities.     
The calculated eclipse light curves show good agreement with observations, with the
changing shape of pre-eclipse and post-eclipse light curves being explained entirely due to 
$\sim 50\%$ variations in the mass accretion rate through the gas stream.
\end{abstract}


\section{Introduction}
\label{intro}

\src~$=$~SDSS~J170213.26+322954.1 is an eclipsing dwarf nova that was classified as an SU UMa cataclysmic variable (CV) ~\citep{2006JBAA..116..187B}. 
The orbital period is about 0.1~day, which in the orbital period gap for CV stars.
The system was extensively observed in quiescence, revealing the strong change in the light curve shape
\cite[][, see Fig.~\ref{fig:data}]{2015ARep...59..288K}.
Orbital light curves demonstrate deep minimum (the orbital phase $\phi_\mathrm{orb}=0$) with a depth of
1.5~mag corresponding to the primary eclipse, and the secondary minimum ($\phi_\mathrm{orb}=0.5$) with
a depth of 0.2~mag.
The most of the orbital light curves (the top and bottom panels in Fig.~\ref{fig:data})
exhibit a 0.2~mag pre-eclipse hump (at $\phi_\mathrm{orb} \simeq -0.2$).
However, at different nights in the quiescent state the source shows different magnitude of the 
plateau between the primary and secondary maxima (at $\phi_\mathrm{orb} = 0.1..0.4$). 
To explain the observed changes in the light curve of \src, \cite{2015ARep...59..288K} proposed a 
phenomenological multi-parametric model of the system including two components 
(an optical star with mass 0.223 $\msun$ filling the Roche lobe, 
a white dwarf with mass 0.91 $\msun$), a 'hot spot' and a 'hot line' 
at the rim of the accretion disc around the white dwarf, with the sizes and temperatures of 
the hot spot and hot line varying in different nights. 

In highly inclined eclipsing binaries the optical light curves are mostly shaped by external parts of
the complicated gas flows around the compact star, and therefore they can be calculated using %simplified 
3D hydrodynamic treatment, with only a few free parameters (primarily, the mass accretion rate supplied from 
the vicinity of the inner Lagrangian point of the binary system). 
In this paper, we perform such 3D hydrodynamical simulations to explain
the observed orbital light curves of \src. The gas-dynamic model includes a cold gas stream which initially 
starts at the inner Lagrangian point and in several orbital periods forms a disc-like structure around the white dwarf. 
To calculate the light curve, we use the cooling function of ionized hydrogen gas. 
% Unlike previous 3D-studies of gas flows in CVs.
%(e.g. \cite{Bisikalo1998} and later papers of this group), 
We take into account partial ionization of hydrogen, which is important from hydrodynamic point of view by contributing to the gas internal energy.
The calculations show that the light curve of \src\ in quiescence is very well described by this model, with only parameter being the mass accretion rate $\dot M$ through the gas stream. Like \cite{1998ARep...42...33B,Bisikalo1998} we found that the pre-eclipse hump can be explained by the existence of 'hot line' in the vicinity of stream and accretion disc intersection. Also, the calculations show absence of the 'hot spot'.

The structure of this paper is as follows. In Section \ref{s:model}, we describe the gas-dynamic model. 
In Section \ref{s:results},  we present the results of 3D numerical modelling. In Section \ref{s:comparison}, a comparison with observations is given and in Section \ref{s:conclusions}, we present our conclusions.

\section{The model}
\label{s:model}

\subsection{Motivation}

\begin{figure}
%\centering
\includegraphics[width=0.475\textwidth]
{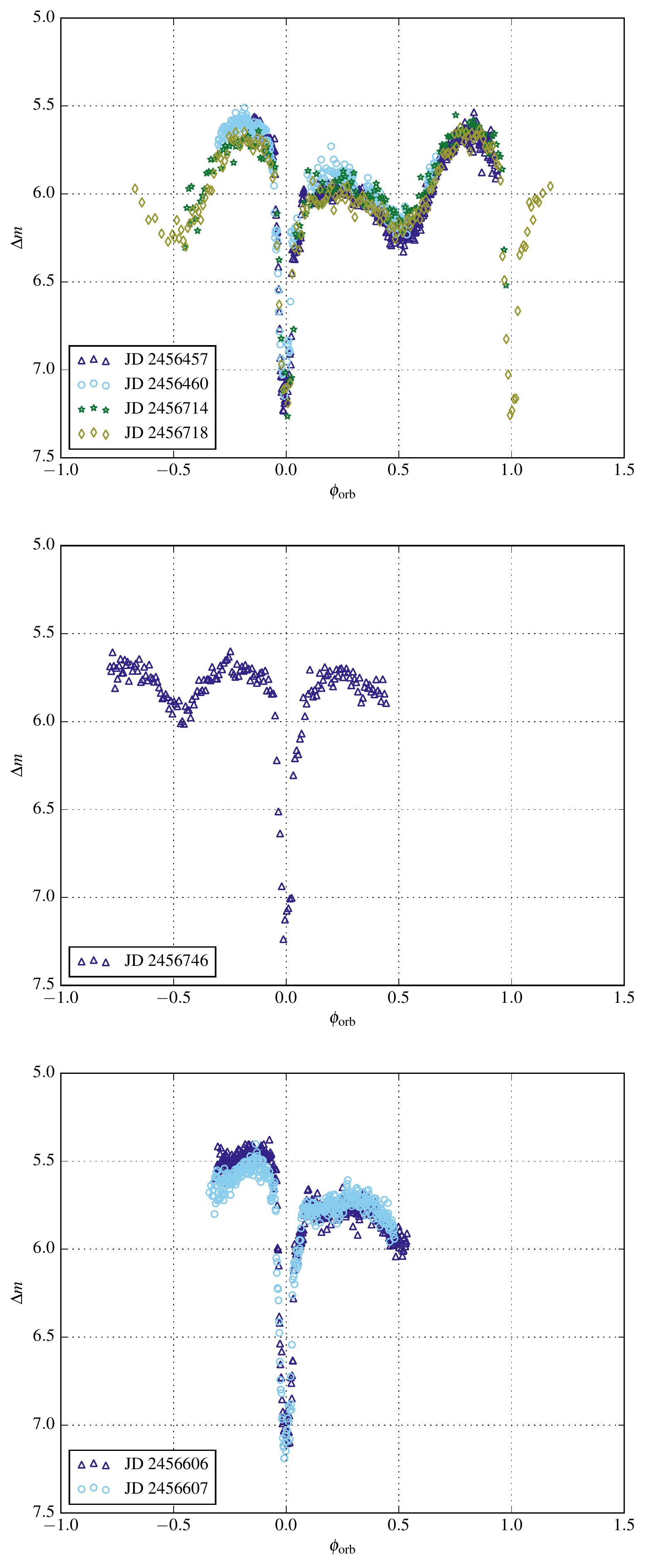}
\caption{Colour online. Observed light curves of V1239 Her in quiescence. Shown is 
difference in white light magnitudes between the comparison star and V1239 Her as a function of 
the orbital phase. 
Top panel: group I light curves (combined JD 2456457, 2456460, 2456714, 2456718 observations);
middle panel: group II light curve (JD 24564746 observations);
bottom panel: group III light curves (combined JD 24564606, 24564607 observations).
Data from \protect\citet{2015ARep...59..288K}. 
\label{fig:data}}
\end{figure}

The aim of this present paper is to provide physical interpretation, by means of 3D-gas dynamic modelling, of 
optical light curves of eclipsing cataclysmic variable V1239 Her in quiescence presented by 
\cite{2015ARep...59..288K}. These light curves obtained without a filter (in white, or integrated, light) are shown in Fig.~\ref{fig:data}. The model includes the gas stream emanating from the inner
Lagrangian point of the Roche lobe filling optical star, which forms a steady-state accretion flow around the compact 
more massive companion (a white dwarf). Our model enables the calculation of large-scale 3D structures in the accretion flow,
which primarily shapes the observed light curve. The gas flow is modelled using  the following assumptions: 
 \begin{itemize}
 \item Partially ionized hydrogen plasma flow is controlled by compressible gas dynamic equations.
 \item No self-gravity and magnetic fields are taken into account.
 \item Plasma is treated as an ideal inviscid gas which heats solely due to conversion of the 
gravitational energy into the internal gas energy.
 \item Energy losses are due to radiation cooling in optically thin collisional plasma (this assumption is
not very adequate but it can be used as first approximation since it strongly facilitates calculations by neglecting radiation transfer equation in simulations).
\end{itemize}
The light curve is  calculated using the radiative cooling function with account for the 3D-structure of the flow integrated
along the line of sight up to an optical depth of $\tau=2/3$.

The observed  light curves of V1239 Her in quiescence, which are 
presented in \cite{2015ARep...59..288K}, can be separated in three groups. 
Group I includes four observations (JD 2456457, 2456460, 2456714, 2456718), which we conventionally will refer to as the 
'regular regime', as shown in Fig.~\ref{fig:data}, top panel. The main feature of group I light curves is the pre-eclipse 
hump (i.e. before the donor star eclipses the accretion disc).
Group III  (Fig.~\ref{fig:data}, bottom panel) differs from group I by general increase in brightness, which is usually related to increase in the mass transfer rate in the binary system.
Group II includes only one observation (Fig.~\ref{fig:data}, middle panel), in which no brightness humps are visible around the eclipse, but the general brightness level is higher than in group I observations, like in group III light curves. 
Note that almost symmetric light curves similar to group II ones are observed on rare occasions,
suggesting a transitional character of these states of the system (see more discussion below). 

The pre-eclipse hump visible in group I and group III light curves of V1239 Her can be related to the presence of a 'hot line' formed in the site of interaction of the gas stream with the outer rim of the accretion disc. To check this hypothesis, we perform adequate hydrodynamical modelling of the accretion disc structure and of the stream-disc interaction region at different mass transfer rates through the gas stream.  

\subsection{Equations and notations}

The mathematical model includes the set of hydrodynamic equations of ideal (inviscid) partially ionized hydrogen gas with
account for the Roche gravitational potential and radiative gas cooling, written in non-inertial reference frame corotating with the binary orbital period:
\begin{gather}
    \frac{\partial \rho}{\partial t} + \vec\nabla \rho \vec{v} = 0, \label{eq:mass}\\ 
    \frac{\partial \rho \vec v}{\partial t} + \vec\nabla\cdot \left(\rho \vec{v}\vec{v} + p \hat{I} \right)= - \rho \vec \nabla \Phi + 2 \rho \vec v \times \vec \Omega, \label{eq:impulse}\\
    \frac{\partial e}{\partial t} + \vec\nabla\cdot \left(\vec v\left(e+p\right) \right)= -\rho \vec \nabla \Phi \cdot \vec v - \Lambda(\rho, i, T), \label{eq:energy}
\end{gather}
where $\rho$ is the density, $\vec v$ is the velocity vector, $p$ is the pressure, $e$ is the total gas energy per unit volume,  
$\vec \Omega$ is the angular rotational velocity of the binary system, 
$\Lambda(\rho, i, T)$ is the radiative gas cooling function,
\begin{equation}
 \Phi = - \frac{GM_1}{\|\vec r- \vec r_1 \|} -  \frac{GM_2}{\|\vec r- \vec r_2 \|} - \frac12 \| \vec \Omega \times \left(\vec r - \vec r_c\right) \|^2,
\end{equation}
$\vec r$ is the radius-vector, $M_1,\, M_2$ are masses of the components, $\vec r_1,\, \vec r_2$ are their radius-vectors, $\vec r_c$ is the system barycentre radius-vector, $G$ 
is the Newton gravity constant. The system of hydrodynamic equations is completed by the equation of state of
partially ionized gas \citep{book_landafshitz_stat1980}:
\begin{gather}
e = \frac32 p + \rho \frac{\Ry}{m_p} i +  \frac{\rho \|\vec v\|^2}{2}, \; p=\rho k T \frac {1+i}{m_p},  \\
i = \left[1 + p \left( \frac{2\piup \hbar^2}{m_e} \right)^{3/2}(k T)^{-5/2} \, \mathrm{e}^{\Ry/kT}   \right]^{-1/2},
\end{gather}
where $i$ is the degree of ionization,  $T$ is the temperature,
$\varepsilon$ is the specific internal energy of gas,
$k$ is the Boltzmann constant,
$\Ry$ is the Rydberg unit of energy, $m_p$ and $m_e$ are the masses of a proton and an electron. 

\subsection{Cooling function}

The integral cooling function for pure hydrogen was calculated as a function of gas
density and temperature assuming local thermodynamical equilibrium (i.e. the Saha-Boltzmann distribution of 
atomic level populations). The cooling function includes radiative transitions between levels, 
two-photon $2s\to 1s$ transition, free-bound and inverse transitions and bremsstrahlung radiation 
 \citep{1993ApJS...88..253S,1985ApJ...297..476B}.
The hydrogen energy levels are taken from the NIST atomic spectra data base, and transitions between the levels 
are calculated according to the Kurucz database  \citep{2002AIPC..636..134K}. 
The two-photon $2s\to1s$ transition emission characteristics are taken from  
\citet{1984A&A...138..495N}, the photoionization cross-sections are calculated in 
\citet{1961ApJS....6..167K}, the mean Gaunt-factor for free-free processes is calculated in 
\citet{1998MNRAS.300..321S}.

\begin{figure}
\centering
\includegraphics[width=0.475\textwidth]{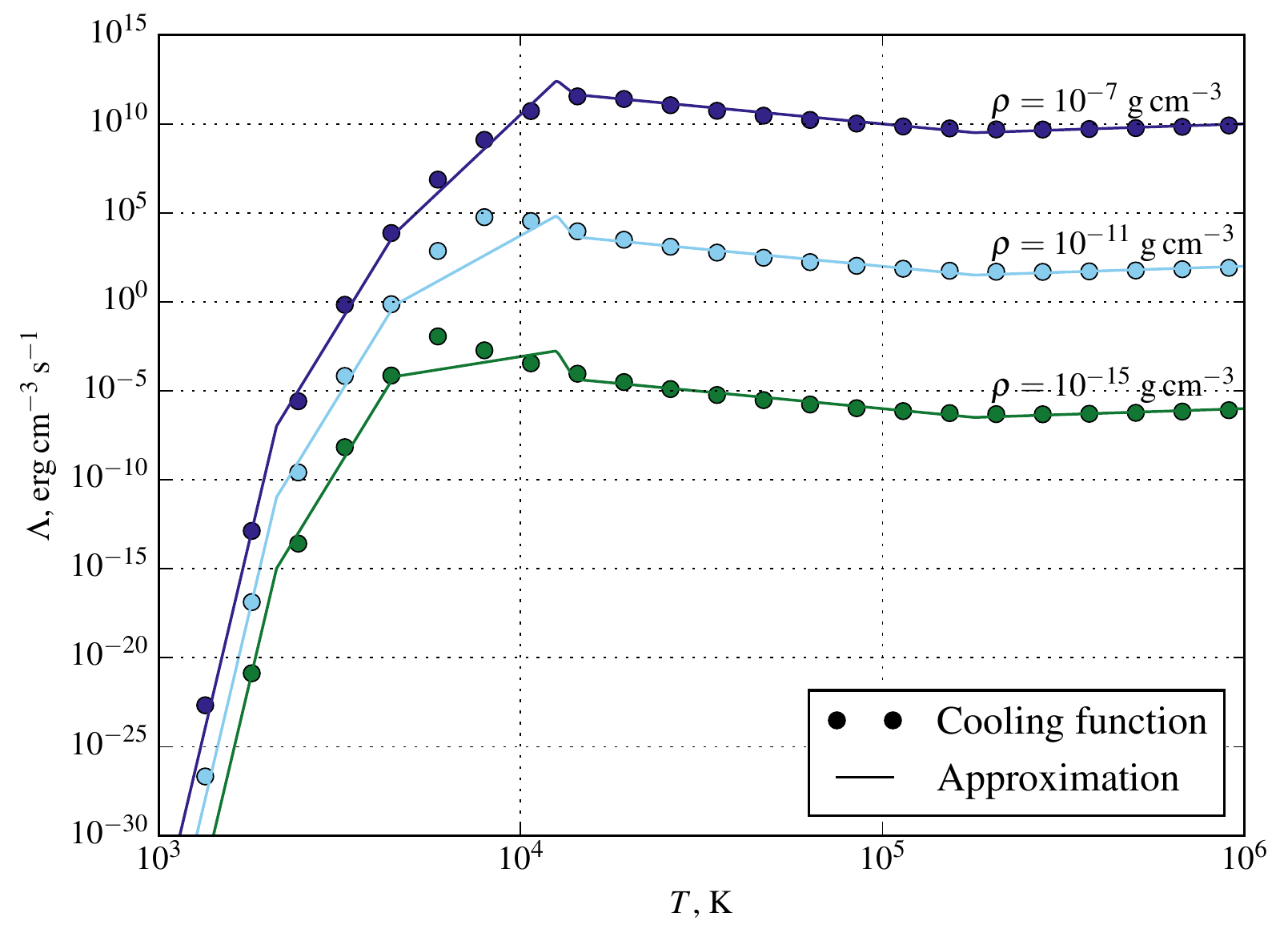}
 \caption{Radiative cooling function and its approximation for different hydrogen gas density.\label{fig:cooling}}
\end{figure}

The radiation cooling function is shown in Fig.~\ref{fig:cooling} and can be written in the approximate form (here $T$ is in K):
%%%
%%%
\begin{gather}
 \lg \Lambda(\rho, i, T) = k(T) \lg \left(i \rho \right) + b(T),\\
 k(T) = \left\{ \begin{array}{cc}
                 1, & \lg T < 3.65, \\
                 2 \lg T-6.3, & 3.65 < \lg T < 4.15, \\
                 2, & 4.15 < \lg T,
                \end{array}
 \right.\\
 b(T) = \left\{  \begin{array}{cc}
                 86.154 \lg T - 286.46, & \lg T < 3.325, \\
                 33.226 \lg T - 110.48, & 3.325 < \lg T < 4.1, \\
                 -1.957 \lg T + 33.77, & 4.1 < \lg T < 5.25, \\
                  0.667 \lg T + 20, & 5.25 < \lg T.
                \end{array}                
 \right.
\end{gather}

\subsection{Dimensionless parameters of the system, initial and boundary conditions}
\label{par:conditions}

The system of hydrodynamic equations was recast to the dimensionless form using the characteristic scales of 
V1239 Her, to wit: 
\begin{itemize}
 \item the length is measured in units of the distance between the components, $L = 0.945 \, \rsun$, where $\rsun$ is the solar radius;
 \item the stellar masses are in units of the total mass of the binary,  $M_0 = M_d + M_a$, where 
$M_d = 0.223\, \msun$ and $M_a = 0.91\, \msun$ are masses of the donor and accretor, respectively, $\msun$ is the solar mass;
 \item the temperature is in units $T_0 = 3500 \;\text{K}$;
 \item the gas density is in units $\rho_0 = 1.6 \cdot 10^{-8} \; \text{g\,cm}^{-3}$.
\end{itemize}
Other scales can be constructed from the basic units in the standard way. The time-scale is in units of the
orbital period of the binary system, $t_0 = 2 \piup \sqrt{\frac{L^3}{G M_0}}$. The angular velocity scale is $\Omega_0 = 1/t_0$, 
and the dimensionless orbital angular velocity of the binary is $\Omega = 2 \piup$.
The cooling function scale is $\Lambda_0 = \rho_0\,L^2\,/\,t_0^3 = 107.5$\,erg\,cm$^{-3}$\,s$^{-1}$.

The origin of a non-inertial reference frame is set at the centre of the donor star, the axis 
$Ox$ passes through the binary components centres. The calculated domain includes a spherical layer centred at 
accretor's centre with the dimensionless inner radius $r_i = 0.01$, which models accretor's surface, and the outer radius
$r_o = 0.64485$. In this setup, the inner Lagrangian point L$_1$ lies at the intersection of the axis  
$Ox$ with the outer boundary of the computational domain.

The initial conditions of the problem describe plasma at rest in the inertial reference frame which rotates around the barycentre 
of the binary system. The plasma density is 
$\rho = 10^{-5}$, the temperature is $T = 10^{-4}$.
These values of the background density and pressure, on the one hand, enables calculations to be performed 
by assuming fluid medium, and on the other hand, allows the gas stream outflow in vacuum to be calculated.

The boundary conditions represent the free outflow at the outer boundary of the computational domain and 
% they are reflecting 
non-reflecting flow at the inner boundary enabling free plasma sinking at the artificial
accretor's surface. In the vicinity of the inner Lagrangian point with the coordinate   
$x=0.35515$, the boundary conditions model the gas stream inflow from the donor star with the density  
$\rho = 1$, temperature $T=1$, and the velocity with non-zero $x$-component only equal to the local speed of sound. The gas inflow
region is taken as a circle with radius  
$r_{jet} = 0.01$ at the external boundary of the computational domain. These parameters give standard mass flux in stream $\dot M_s \approx 4.9 \cdot 10^{16}\,\mathrm{g}\,\mathrm{s}^{-1}  \approx 7 \cdot 10^{-10}\,\msun\,\mathrm{yr}^{-1}$.

\subsection{Numerical scheme}

To model the plasma flow, an explicit difference numerical scheme of the Godunov type is constructed using 
tetrahedron unstructured grids. The scheme is additive and employs splitting of physical processes -- when passing from 
a given time layer to the next one, the convective terms in equations 
(\ref{eq:mass})--(\ref{eq:energy}) are first integrated, and then the gravitational source term in equations 
(\ref{eq:impulse})--(\ref{eq:energy}) and the sink radiation function (equation \ref{eq:energy}) are integrated. 
Importantly, the radiative plasma cooling occurs on much shorter time-scale than gas dynamic processes. Therefore, 
when integrating the cooling function in passing from one time layer to the next one, the time step is split in substeps. 
  
The numerical flux at the tetrahedron faces is taken in the form of a modified HLLC (Harten --- Lax --- van Leer --- Contact) flux \citep{Toro1994}
specified for the sound speed calculation in partially ionized plasma according to \citet{book_cox_and_giulis}:
\begin{equation}
 \vs =\sqrt{ \frac{p}{\rho} \frac{ 5 + i(1-i) \left( \frac52 + \Ry \right)^2 }{ 3 + i(1-i) \left[ \left( \frac32 + \Ry \right)^2 + \frac32 \right] }}. \label{eq:soundSpeed}
\end{equation}
Note the similarity of this approach to the modelling of a perfect gas flow. In the last case, the sound speed is calculated as 
\begin{equation}
\vs = \sqrt{\gamma \frac p \rho}, \label{eq:perfectSoundSpeed}
\end{equation}
where $\gamma$ is the adiabatic index. For the ionization degree
$i = 0$ or 1 formula (\ref{eq:soundSpeed}) 
gives the same values as (\ref{eq:perfectSoundSpeed}) 
for monoatomic gas with $\gamma = 5/3$. 
For intermediate values of the ionization degree $i$, 
the coefficient in equation (\ref{eq:soundSpeed}) turns out to be less than $5/3$, 
which can be interpreted as the partially ionized plasma being more compressible than a monoatomic perfect gas.    

The time step of integration of the convective part of the system 
(\ref{eq:mass})--(\ref{eq:energy}) was calculated dynamically with the Courant number set to 
$C = 0.4$. 
The time step splitting in integration of the radiative cooling function was also performed dynamically, 
and the number of substeps was at least $10$.

Complex and labour consuming calculations to model the accretion flow dynamics forced us to elaborate
a dedicated parallel code for computer cluster systems based on {\sc MPI} technology. The code is based on splitting of the computational domain
into subdomains in which the solution is calculated in parallel.
Test parallel calculations demonstrated that the acceleration factor is $13.2$ for calculations using $12$ and $192$ processing cores, which is about $82.5\%$ of the maximum attainable for the linear acceleration.

\subsection{Calculating the light curve}

The total energy flux $F$ observed from the model system is calculated by integrating the visible structure along observer's line of sight:
\begin{equation}
    F = \int_S{ I(x, y) \, \frac{\di S}{d^2} }\,,
\label{eq:total_flux}
\end{equation}
where $I(x, y)$ is the intensity in the plane perpendicular to the line of sight (picture plane), $x$ and $y$ are the coordinates on this plane, $d$ is the distance to the source.
The intensity $I(x,y)$ is calculated as the cooling function $\Lambda$ averaged over the hydrodynamical time step at the optical depth $2/3$:
\begin{equation}
\begin{split}
 I(x,y) &= \frac{\bar{\Lambda}(x,y,z^*)}{4\piup},\\ 
 z^*: \tau(x,y,z^*) &= \int_0^{z^*}{ \alpha_\mathrm{Pl}(x,y,z') \, \di z' = \frac 23 },
\end{split} 
 \label{eq:intensity}
\end{equation}
% The intensity $I(x,y)$ is obtained from the radiation transfer equation:
% \begin{equation}
%  \begin{split}
%     I(x,y) &= \int_0^\infty{ \frac{\bar{\Lambda}(x,y,z)}{4\,\piup} \, e^{-\tau(x,y,z)} \, \di z } + \frac{\sigma_\mathrm{SB} T_\star^4}{\piup} \, e^{- \tau_\star} \,\\
%     \tau(x,y,z) &= \int_0^{z}{ \alpha_\mathrm{Pl}(x,y,z') \, \di z' },
%  \end{split}
% \label{eq:intensity}
% \end{equation}
where $z$ is the distance along the line of sight, 
% $\bar{\Lambda}$ is the cooling function $\Lambda$ averaged over the hydrodynamical time step, 
$\tau$ is the optical depth between the observer and the volume element, 
$\alpha_\mathrm{Pl}$ is the Planck mean absorption coefficient. For the donor star, constant surface temperature was used to obtain its contribution to the observed flux.
%, and index $\star$ marks the donor star surface.

The absorption coefficient $\alpha_{Pl}$ is set to infinity in the donor star and is the Planck mean elsewhere.
The Planck mean absorption coefficient is defined as
\begin{equation}
    \alpha_\mathrm{Pl}(\rho, T) = \frac{ \int_0^\infty{\alpha_\nu(\rho,T,\nu) \, B_\nu(T,\nu) \, \di\nu} }{ \int_0^\infty{B_\nu(T,\nu) \, \di\nu} },
\label{eq:planckmean_absorption}
\end{equation}
where $\alpha_\nu$ is the spectral absorption coefficient and $B_\nu$ is the Planck function.
As the radiation cooling time in our calculations is much shorter than the hydrodynamical one, 
we assume radiative equilibrium and can use the integrated Kirchhoff's law:
\begin{equation}
    \int_0^\infty{\alpha_\nu(\rho,t,\nu) \, B_\nu(T,\nu) \, \di\nu } = \int_0^\infty{ \frac{\Lambda_\nu}{4\,\piup} \, \di\nu } = \frac{ \Lambda} {4\,\piup},
\label{eq:krchhoffs_law}
\end{equation}
where $\Lambda_\nu$ is the spectral cooling function that differs from the spectral emission coefficient by a factor of $4\,\piup$.

Substituting the last formula into equation~\eqref{eq:planckmean_absorption} and using the 
integral Planck function $\int_0^\infty{B_\nu(T, \nu) \di\nu} = \sigma_\mathrm{SB} T^4 / \piup$, 
we obtain the relation between the absorption coefficient and the cooling function:
\begin{equation}
    \alpha_{Pl}(\rho, T) = \frac{\Lambda(\rho, T)}{4 \sigma_\mathrm{SB} T^4},
\label{eq:alpha_Pl}
\end{equation}
where $\sigma_\mathrm{SB}$ is the Stefan--Boltzmann constant.

\subsection{Simulation setup}
The computational domain was taken in the form of a sphere with dimensionless radius $r_o=0.64485$ with 
whole sphere of radius $r_i=0.01$ modelling the accretor. The computational domain includes the accretor's Roche lobe
and touches the donor's Roche lobe at the inner Lagrangian point L$_1$ from which the matter is ejected towards the accretor. 
The calculations are carried out using a tetrahedral grid with condensation near the equatorial plane of the accretion disc.
The approximate grid step at the accretor's surface is $0.0025$, near the disc-stream interaction region is $0.02$, and at the maximum distance from the accretor is $0.05$. The grid includes $560,000$ cells. The calculation time interval is equal to $26$ orbital periods splitting into $2,300,000$ time-steps. 

To construct the light curve, the picture plane $2.5$ (in the equatorial plane)  by $1$ (along the vertical axis) 
was used. The picture plane is located at a distance of $1.5$ from the system's barycentre, is inclined by $85^\circ$ to the 
orbital rotation axis and rotates in the azimuthal direction according to light curve phase. For each phase, the brightness distribution in the picture plane is calculated, and the integral brightness corresponds to one point on the light curve.
The light curve includes $100$ equidistant phases. 

The picture plane is split into $750\times300$ ($225,000$) pixels, and tracing lines are constructed along the pixel normals.   
Each tracing line is split into $3,000$ intervals, in which the optical depth was numerically integrated. In addition to the computational domain with the disc, a model of the donor star is calculated. The donor star is assumed to fill its Roche lobe and has a constant temperature and optical depth much larger than unity.
The length of each tracing line was taken to be equal to $3$, enabling the optical depth integration through the entire model of the binary system including the computational domain and the donor star. In the optical depth calculations, the integration started from the picture plane along the tracing line up to reaching the optical depth $2/3$\footnote{Note that this value of the optical depth is taken rather arbitrarily; the use of other value (say, $1$) does not affect the results.}. The brightness in each pixel of the picture plane was taken to be equal to either the mean value (over the time step) of the cooling function in the corresponding cell of the computational domain of the disc, or the donor star surface brightness, or zero if the optical depth $2/3$ was not attained. 
Next, all pixel brightnesses from the picture plane were sum up to obtain one point on the light curve corresponding to the
given phase of the picture plane location.

\section{Results}
\label{s:results}

\begin{figure}
\centering
\includegraphics[width=0.475\textwidth]{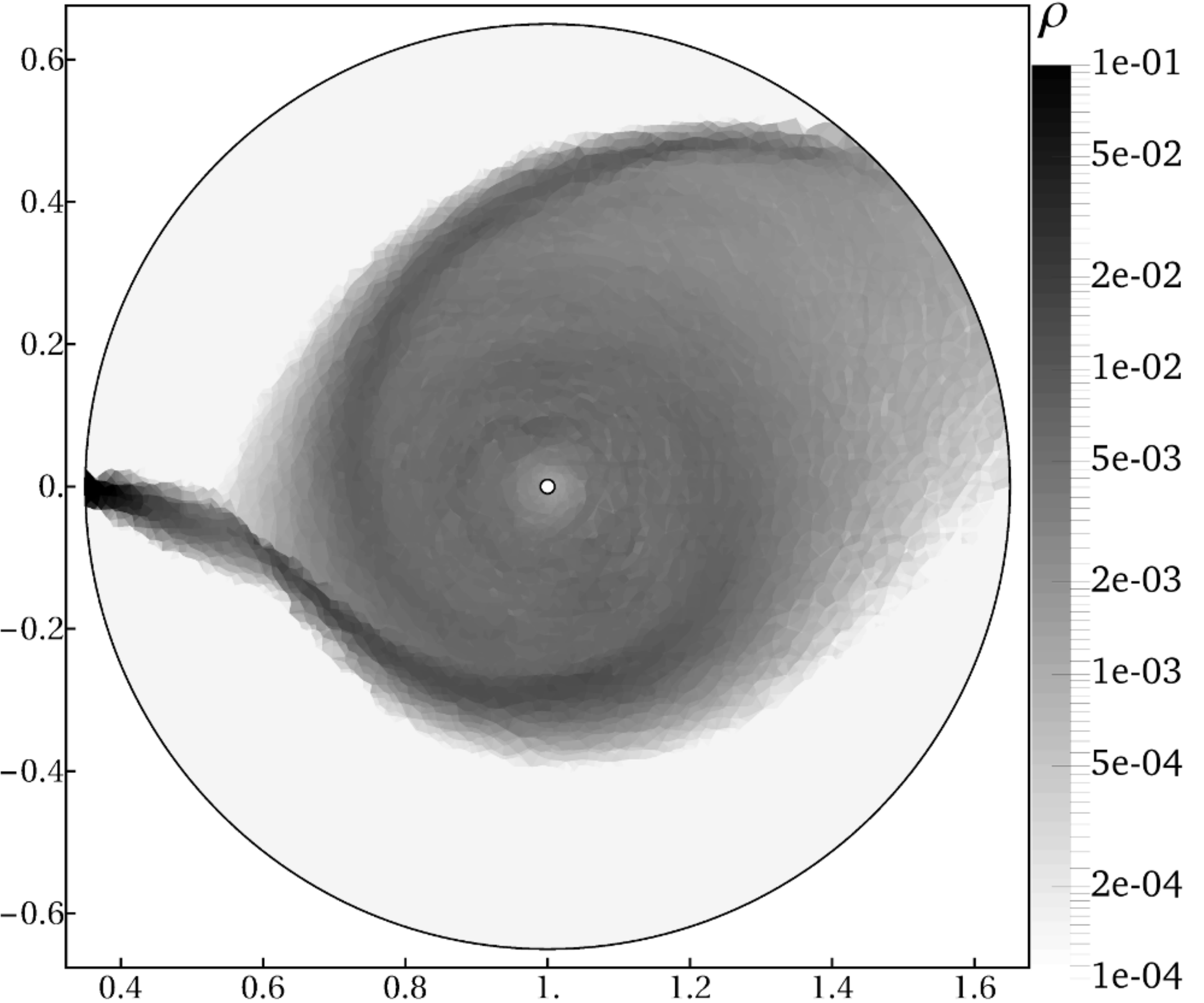}
% \vfill
\includegraphics[width=0.475\textwidth]{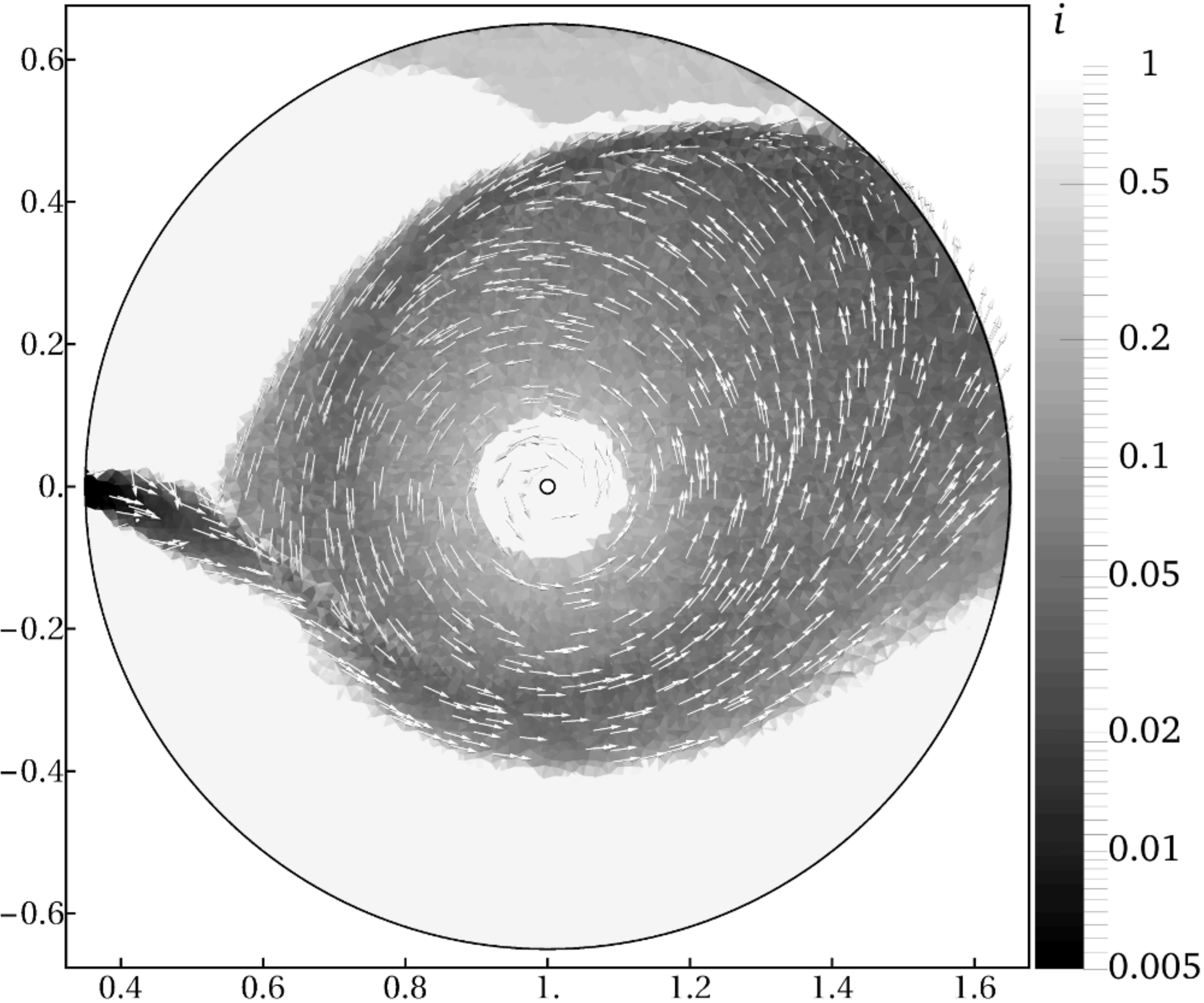}
\includegraphics[width=0.475\textwidth]{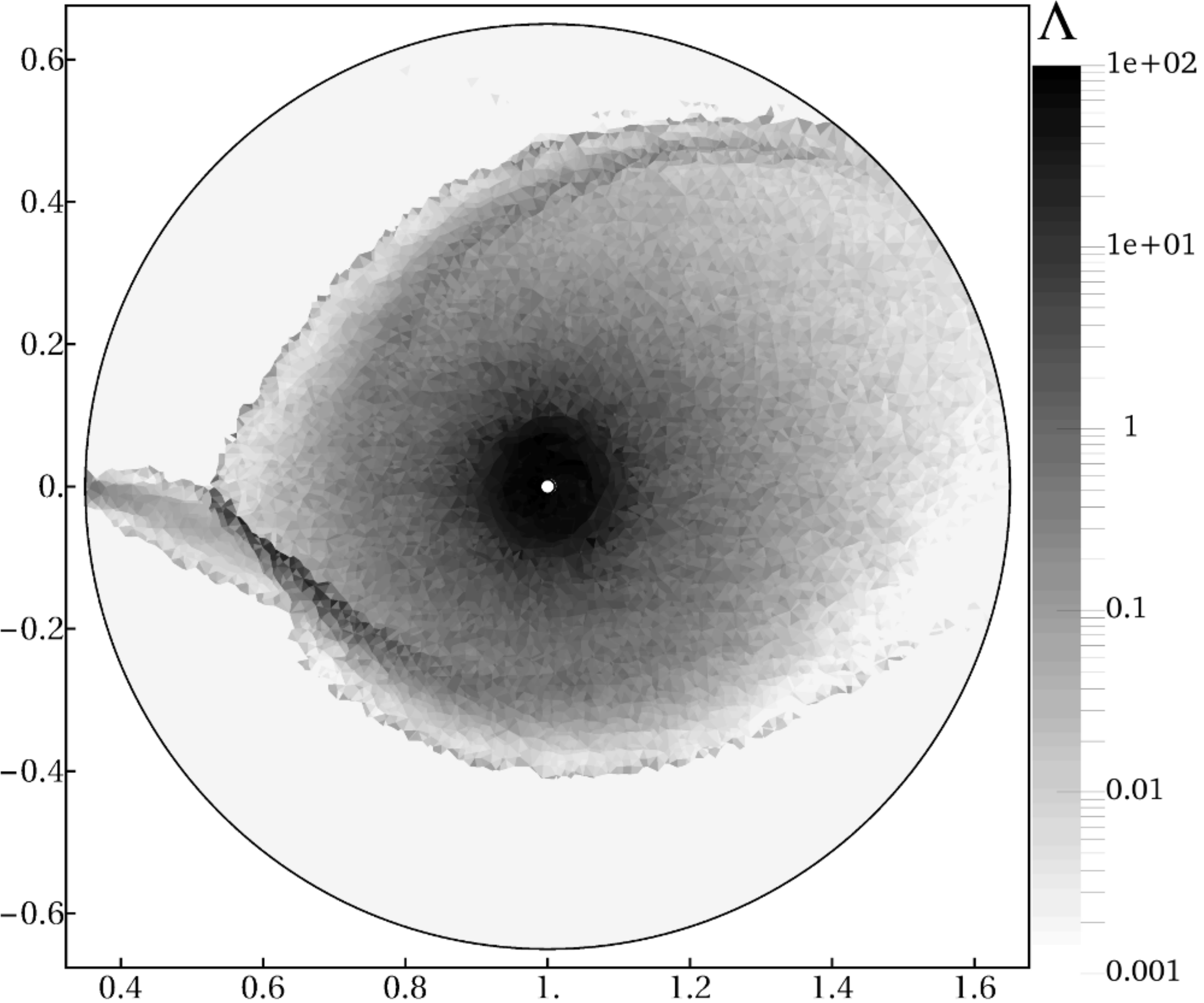}
\caption{Face-on view of the accretion flow in V1239 Her with accretion rate $\dot M_s$. Top panel: Gas density. 
Middle panel: Ionization degree (grey colour) and velocity (white arrows) field. 
Bottom panel: Volume emissivity (radiative cooling function) $\Lambda$.
\label{fig:faceon}}
\end{figure}

Here we present the results of 3D modelling of the accretion flow on to the gravitating centre
in a binary system with orbital parameters of V1239 Her. First, in Fig.~\ref{fig:faceon} we 
show the face-on view of the flow taken after $17$ orbital periods since the start of the flow 
from the vicinity of the inner Lagrangian point L$_1$. In the top panel, grey colour shows the density distribution.
An extended density enhancement at the interaction region of the gas stream with 
disc-like flow is clearly visible. Also, one can see the spiral density wave starting on the front (relatively to system rotation) edge of the disc
and twisting to the accretor. Similar spiral structures have been found in other numerical simulations of accretion discs (see, for example, \citet{2016ApJ...823...81J}) and can be additional sources of the disc heating and angular momentum transfer. The top part of the spiral wave is the shock wave at the interface between accretion disc and background medium.

In the middle panel, grey colour shows the ionization degree and arrows show the projection of the velocity direction of randomly taken particles on the orbital plane. 
In the inner part of the flow, the particles move along almost Keplerian trajectories.
The difference of the particle velocities from Keplerian ones is due to taking into account 
the pressure gradient when calculating particle trajectories, as well as the presence of a (small) numerical viscosity, which is inherent in the HLLC method. 
Also, note that part of the injected gas particles leaves the computational domain through the vicinity of the L$_2$ point (to the right), so that only 
about $50$~per cent of the injected mass reaches the inner boundary (the white central circle).  

\begin{figure}
\centering
\includegraphics[width=0.475\textwidth]
{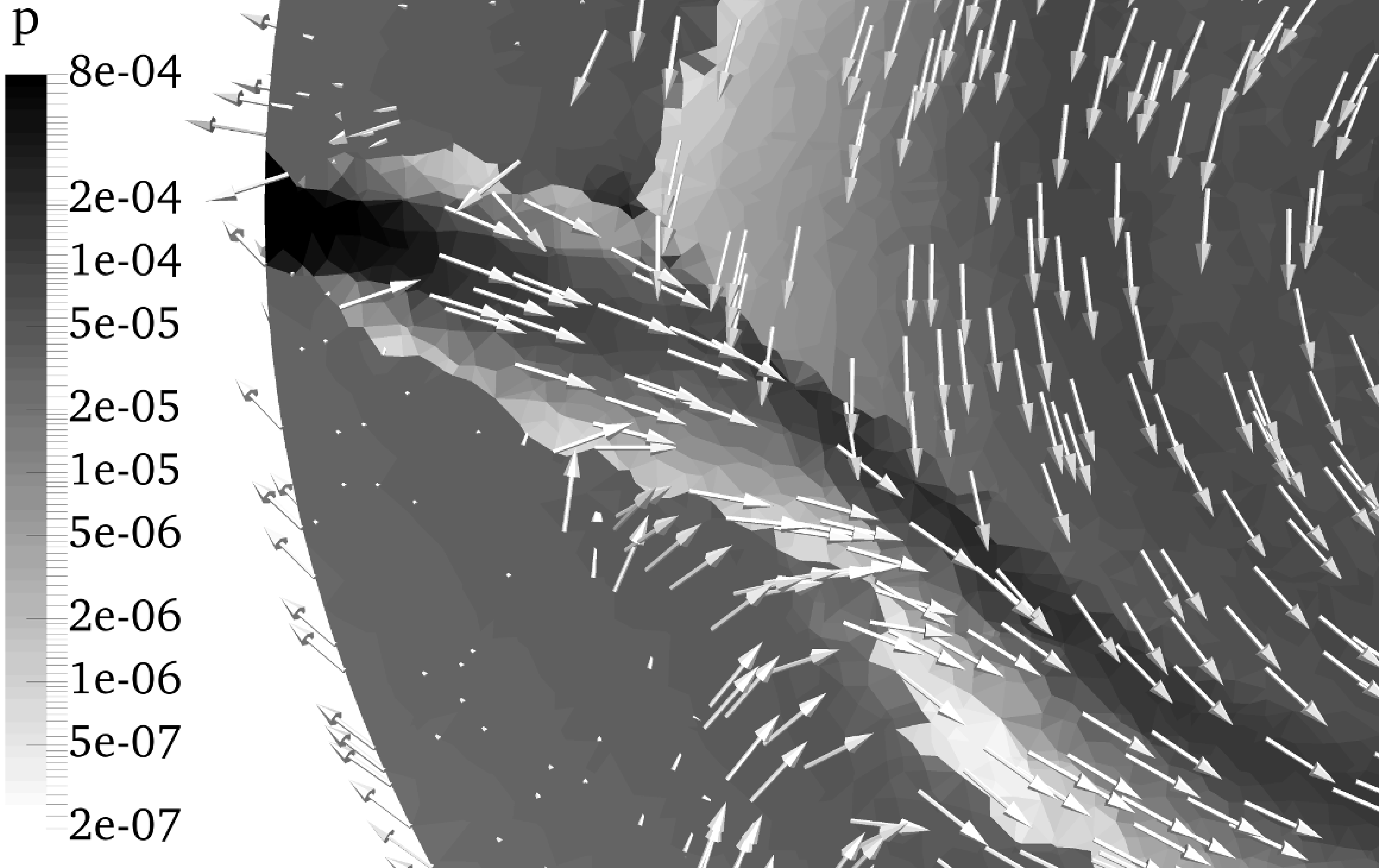}
\caption{Shock wave at the stream-disc interaction region. Pressure is in colour, the instant velocity vectors are shown by arrows with the length of vectors being proportional to the velocity modulus. The mass inflow through the discontinuity is seen, indicating the presence of a shock wave.\label{fig:blast}}
\end{figure}

The bottom panel of Fig.~\ref{fig:faceon} presents the cooling function $\Lambda$ of gas in 
units of $\Lambda_0$. It is seen that the maximum energy output occurs in the disc central parts adjacent to the accretor. 
At the same time, a lot of energy is also released in the interaction region between the stream and the disc rim. 
Here the gas density and temperature increase occurs because the stream flow crosses the transversal flow at the disc rim. 
As the gas density in the stream is much higher than at the disc rim, the stream motion dominates the flow structure, 
and the disc matter crosses the shock wave at the disc boundary to become involved by the stream into the next turn. 
When crossing the shock wave, the disc matter density increases and form the emitting region that is sometimes referred to as the 
'hot line' \cite{Bisikalo2000}. The shock is zoomed in Fig.~\ref{fig:blast}. In the shock region gas is predominantly ionized, 
which increases the radiation energy output. As the Planckian optical depth is proportional to the cooling function (see Eq.~(\ref{eq:alpha_Pl})),
the bottom panel also reflects the optical depth perpendicular to the disc plane. The typical Planckian optical depth turns out to be around 7,
suggesting the need for radiation transfer calculation (the Rosseland optical depth, however, remains much smaller). However, when observing the disc edge-on
(as is the case for V1239 Her), for calculations of the emergent radiation only uppermost optically thin layers are important.

Note that the gas is hot and fully ionized immediately close to the accretor, in the stream-disc interaction region, as well as in a rarefied disc envelope, while the stream gas is initially neutral and is ionized by approaching the disc. 
The use of the model of partially ionized gas enables the formation of a more condensed and, hence, more bright, as well as less hidden from the observer 'hot line'. The latter means that the optical depth of the gas along the line of sight from the outer part of the disc must be much smaller than that obtained from the gas density distribution in calculations by \cite{Bisikalo2000}. This allows us to directly calculate the light curves from the hydrodynamic modelling using the emitted energy and the optical depth along the line of sight at any point of the picture plane.  

\begin{figure}
%\centering
\includegraphics[width=0.475\textwidth]{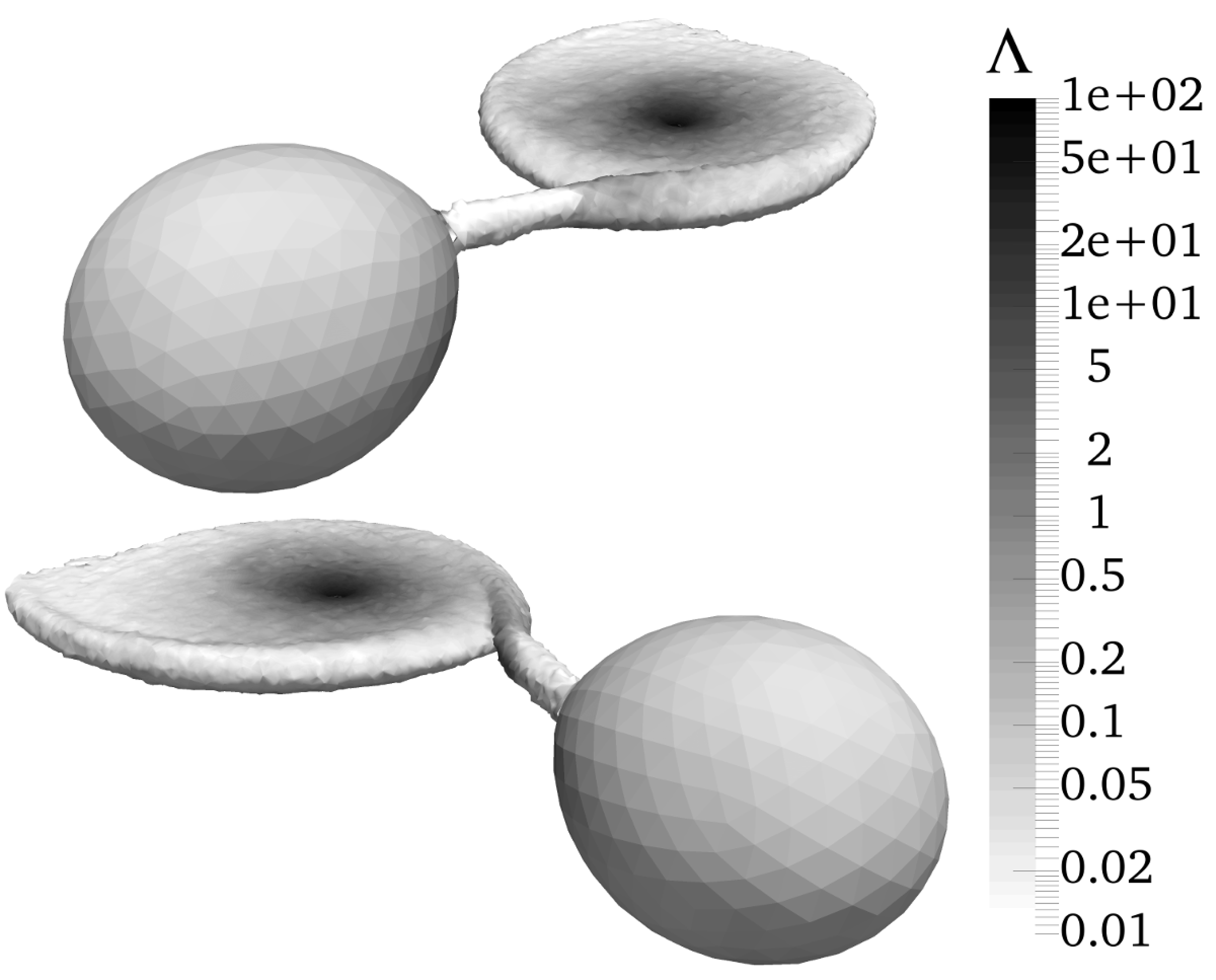}
\caption{3D view of the system before (top) and after (bottom) eclipse.
Shown is the surface of constant density $\rho=0.001$. Gradations of grey colour are used for the volume emissivity $\Lambda$ in the units of $\Lambda_0$.
\label{fig:3Dview}}
\end{figure}

The disc structure of the flow (Fig.~\ref{fig:3Dview}) is formed quite rapidly, during 2-3 orbital periods after the beginning of gas outflow from the L$_1$ point region. At the same time, details of the flow structure, which affect the amount of the energy output and, hence, the light curve shape, require a longer time, of the order of dozen orbital periods after the outflow beginning and the disc formation, and 4-5 orbital periods after the accretion rate change.  

% with observations was also obtained for the transitional flow regime (group III light curves), which is represented by only one observed light curve. The coincidence of the observations with the numerical model results sugges

\section{Comparison with observations}
\label{s:comparison}

\begin{figure}
\centering
\includegraphics[width=0.4\textwidth]{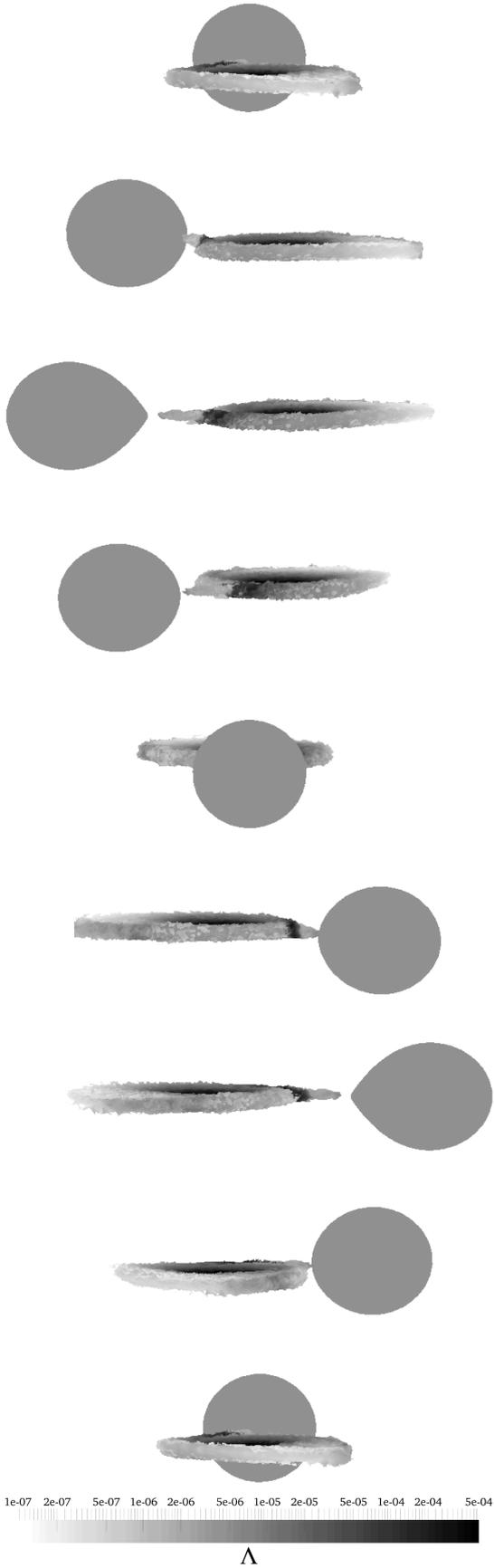}
\caption{Colour animation is available online. View for observer of the system for inclination $85^\circ$ at orbital phases $\phi_\mathrm{orb}$ = $-$0.5;\; $-$0.375;\; $-$0.25;\; $-$0.125;\; 0;\; 0.125;\; 0.25;\; 0.375;\; 0.5. Accretion rate is $\dot M_s$.
\label{fig:phases}}
\end{figure}

The results of hydrodynamic modelling were used to construct light curves of the binary system 
V1239 Her, as well as to calculate the system brightness distribution in the picture plane at different orbital phases.
The brightness distribution at orbital phases 
$\phi_\mathrm{orb} = -0.5;\; -0.375;\; -0.25;\; -0.125;\; 0;\; 0.125;\; 0.25;\; 0.375;\; 0.5$ 
for the binary orbital inclination angle 
$i = 85^\circ$ are shown in Fig.~\ref{fig:phases}.

\begin{figure}
\includegraphics[width=0.475\textwidth]{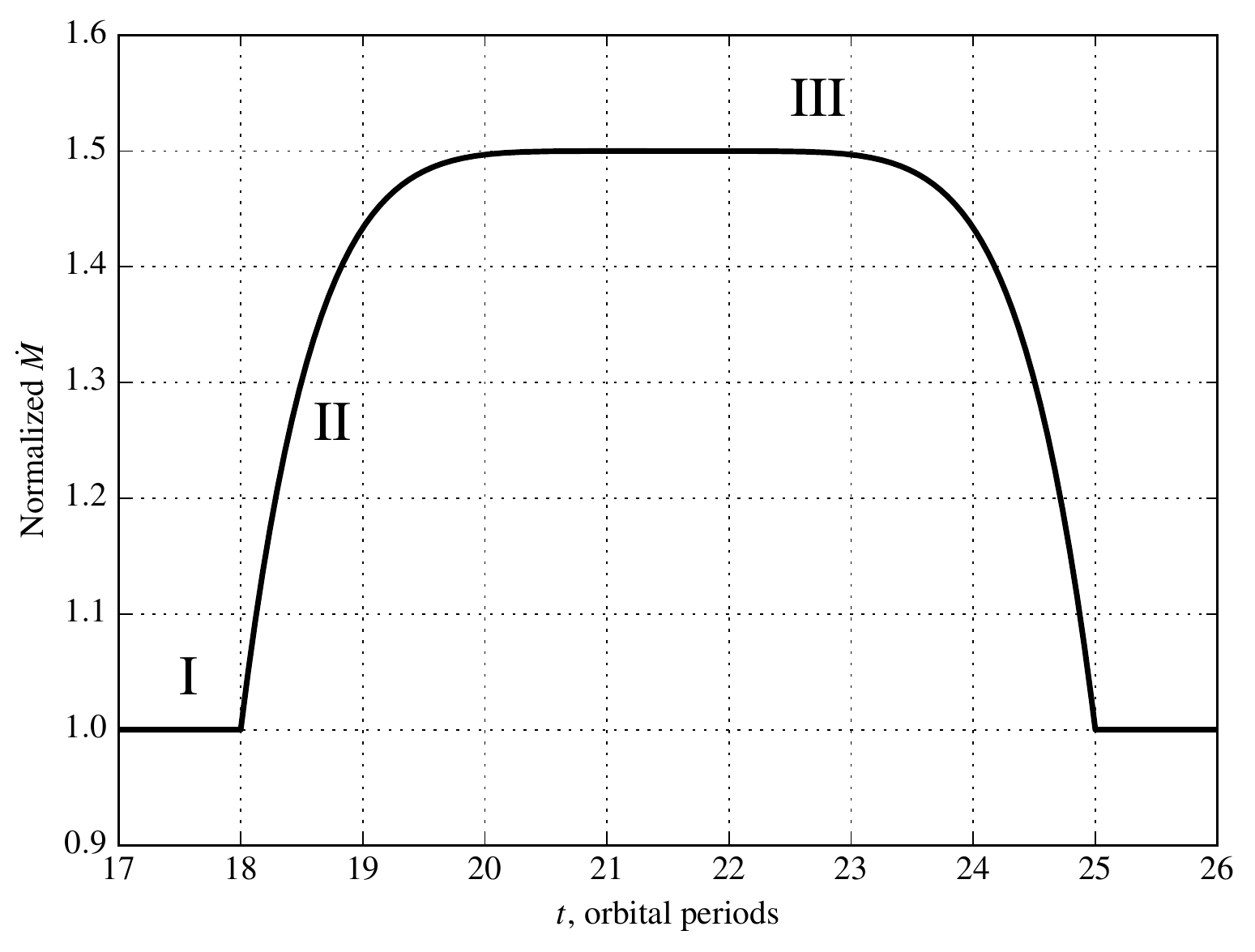}
\caption{The accretion rate variation since 17 until 26 orbital period in units of $\dot M_s$. \label{fig:accRate}} 
\end{figure}

To model light curves for different observation groups discussed above, calculations with variable mass outflow rates from the vicinity of the L$_1$ point were performed. The increase in the system brightness is related to enhanced 
mass accretion rate and, therefore, to the density increase both in the central region around the accretor and in the 'hot line' region. The calculations showed that system's brightness responds to the accretion mass rate growth quite rapidly, within one orbital period. Therefore, to model in one run group I, II and III light curves, the accretion rate was changed as follows: 
\begin{enumerate}
 \item since time $t=0$ until 18 the initial accretion rate was assumed to be constant and was equal to $\dot M_s$;
 \item since $t=18$ until 25 the accretion rate increased by $1.5$ times as shown in Fig.~\ref{fig:accRate};
 \item after $t=25$ the accretion rate is set to the initial value $\dot M_s$.
\end{enumerate}

\begin{figure}
\centering
\includegraphics[width=0.475\textwidth]
{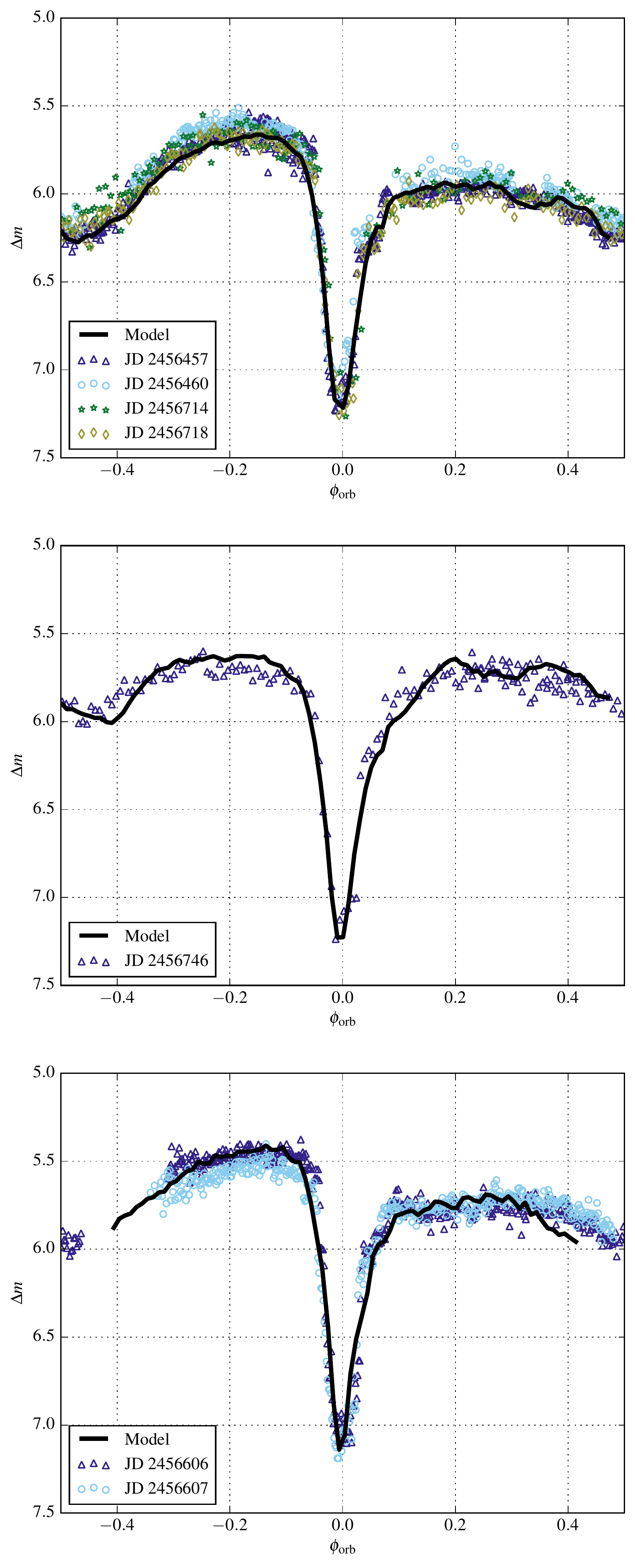}
\caption{Colour online. Calculated (black line) and observed (dots) 
group I, II, III light curves of V1239 Her from Fig.~\ref{fig:data} (top, 
middle and bottom panels, respectively). Accretion rate of the steady-state group I and III (top and bottom panels, respectively) differs by 50\%. 
The intermediate time light curve (group II, middle panel) corresponds to the transition state.
\label{fig:theor_lc}}
\end{figure}

Figure~\ref{fig:theor_lc} shows the model light curves obtained by integrating the cooling function according to 
equation (\ref{eq:intensity}) on the picture plane over the 18th orbital period for group I light curves (the initial mass accretion rate), over the 23d orbital period for group II light curves (for the enhanced mass accretion rate) and over the 19th orbital period for group III light curves (the transitional regime).

The calculation demonstrated that the general brightness level of the system is determined by the visible 
area of the intensively radiating, hot and dense region around the accretor, whereas the detailed structure of the
light curve is determined by the viewing angle of the 'hot line'. More precisely:
\begin{enumerate}
 \item The presence and the shape of the pre-eclipse brightness hump is due to the increase of the visible area of the hot line.   
 \item The formation of the brightness hump starts at the orbital phase 
$\phi_{orb} = -0.4$, and the brightness grows with increasing visible area of the hot line.
 \item The presence of a 'shelf-like' feature at the phase $\phi_{orb} \approx 0.075$ 
is explained by the going out of the eclipse by the donor star of the hot line, which is end viewed at this phase. 
 \item At later phases the hot line goes off the line of sight and is hidden by the optically thick disc, which manifests as the brightness decrease at phases $\phi_{orb} = 0.3..0.35$.
 \item At phases below $-0.4$ and above $0.35$, the hot line on the disc rim is not seen, and at these phases  system's brightness is minimum (apart from the eclipse at the phase $\phi_{orb} = 0$).
\end{enumerate}

Good agreement of the numerical model with observations was also obtained for the transitional flow regime (group II light curves), which is represented by only one observed light curve.
The fit of the numerical model to observations suggests that the corresponding light curve was observed during a close to linear increase in the mass accretion rate.
The duration of the mass accretion rate growth itself is close to $0.5-1$ orbital period, and prior to the eclipse  ($\phi_\mathrm{orb} = 0$) the system has no time to 'feel' the increasing mass accretion rate, so that the pre-eclipse brightness hump does not exceed that of group I regular regime.
After the eclipse, the general increased energy release in the accretion disc and (predominantly) in the 'hot line' is already established, and the brightness level after the eclipse turns out to be comparable to the pre-eclipse one, which explains the appearance of a nearly symmetric light curve. 

The deviation of the calculated light curve from group III observations is related to the fact that the disc is not fully relaxed after four orbital periods since the accretion rate growth. In particular, this is supported by visible light curve oscillations after the eclipse seen at orbital phases 
$\phi_\mathrm{orb}=0.2-0.4$ and the brightness decrease after the phase $\phi_\mathrm{orb} = 0.4$.
Apparently, the observational data correspond to a steady-state flow at the increased mass accretion rate. Generally, good fit of the model to observations at two regular flow regimes in the system can be noted. 
 
Apparently, the rate of gas outflow from the vicinity of the inner Lagrangian point L$_1$ is variable and can relatively insignificantly change within one orbital period (which corresponds to quite noticeable brightness dispersion seen on group I and III light curves). On rare occasions, it can strongly vary leading to transformations of the accretion flow in the disc with ultimate establishing of a new regular regime. The presence of a visible pre-eclipse brightness hump suggests a steady-state flow, while its absence or low value may indicate the transitional character of the flow. Apparently, variations in the mass accretion rate through the vicinity of the inner Lagrangian point are related to physical processes in the donor star and may be due to irregularity in the illumination of its surface by hard emission from the central parts of the accretion disc (the reflection effect).    

\section{Conclusions}
\label{s:conclusions}

In this paper, we presented a 3D time-dependent gas-dynamic model of gas flows around a compact star in close binaries system, 
in which the optical star fills its Roche lobe. Initially, the cold gas stream flows from the 
vicinity of the inner Lagrangian point and in several orbital periods forms a quasi-steady disc-like  
structure around the compact object. The model takes into account gas cooling and partial ionization of hydrogen. 
The model was applied to the eclipsing cataclysmic variable, V1239 Her, which is a SU UMa CV falling within the CV period gap. 
We find that the model light curve describes well the ones of V1239 Her in quiescence, with the 
observed changes in the light curve shape being solely due to changing mass accretion rate through the gas stream. 
We discuss in detail the formation of the apparent features in the light curve (pre-eclipse brightness hump and
post-eclipse behaviour) and come to the conclusion that they can be explained by $\sim 50\%$ changing mass accretion rate in the system.

\section{Acknowledgements}
This work is supported by the Russian Science Foundation grant 14-12-00146.
We used the computation facilities supported by M.V.~Lomonosov Moscow State University Programme of Development
and the hybrid calculation cluster K-100 of the M.V.~Keldysh Institute of Applied Mathematics RAS.

\bibliographystyle{mnras}
%\expandafter\ifx\csname natexlab\endcsname\relax\def\natexlab#1{#1}\fi
\bibliography{v1239her}

% Don't change these lines
\bsp	% typesetting comment
\label{lastpage}
\end{document}